\documentclass[final,1p,times]{elsarticle}

\usepackage{sidecap}

\usepackage{amssymb}

\usepackage{lineno}

\biboptions{sort}

\biboptions{sort&compress}

\journal{Nuclear Physics A}

\begin{document}

\begin{frontmatter}



\title{Isospin Decomposition of the Photoproduced $\Sigma\pi$ 
System Near the $\Lambda(1405)$}


\author{R. A. Schumacher\corref{cor1}} 
\cortext[cor1]{Corresponding author: schumacher@cmu.edu}
\address{Department of Physics, Carnegie Mellon University, Pittsburgh, PA 15213, USA}

\author{K. Moriya}
\address{Department of Physics, Indiana University, Bloomington, IN 47405, USA}

\begin{abstract}
Recent experimental results for the reaction $\gamma + p \to K^+ +
\Sigma + \pi$ from CLAS at Jefferson Lab are discussed. It was found
that the mass distributions or ``line shapes'' of the three charge
combinations $\Sigma^+ \pi^-$, $\Sigma^0 \pi^0$ and $\Sigma^- \pi^+$
differ significantly.  Our results show that the $\Lambda(1405)$, as
the $I=0$ constituent of the reaction, must be accompanied by an $I >
0$ component.  We discuss phenomenological fits to the data to test
the possible forms and magnitudes of these amplitudes.  A
two-amplitude $I=0$ fit of Breit-Wigner form to the $\Sigma^0\pi^0$
channel alone works quite well.  The addition of a single $I=1$
amplitude works fairly well to model all the line shapes
simultaneously.

\end{abstract}

\end{frontmatter}



\section{Introduction}
\label{sec:introduction}
The CLAS Collaboration at Jefferson Lab has measured~\cite{KMRS}, as
part of a large program of photoproduction reactions in the GeV energy
region, the channel $\gamma + p \to K^+ + \Sigma + \pi$.  The
$\Sigma\pi$ invariant masses range was from very close to threshold at
1328 MeV to well beyond 1520 MeV where the $\Lambda(1520)$ is
prominently produced.  In between these limits, the $\Lambda(1405)$ is
produced, as well as the $\Sigma(1385)$ and perhaps other possible
states.  The CLAS experiment is the first to simultaneously map out
this mass range in all three charge combinations $\Sigma^+\pi^-$,
$\Sigma^0\pi^0$ and $\Sigma^-\pi^+$.  This is interesting because
comparison can reveal (1) whether there is substantial contribution of
$I>0$ amplitudes to the reaction mechanism even after the
$\Sigma(1385)$ is removed, and (2) how strong the channel coupling is
between the $\Sigma\pi$ final state and the competing $N\bar{K}$ final
state.  Thus, while there have been hints of line shape distortions of
the $\Lambda(1405)$ in earlier experiments~\cite{Hemingway,Thomas,
Braun:1977wd,Ahn,Niiyama,Zychor,Agakishiev:2012xk} the CLAS experiment
offers the best look so far at this phenomenon.

\begin{figure}
\centering
\begin{minipage}[]{0.58\linewidth}
  \vspace{-1.0cm}
  \resizebox{1.0\textwidth}{!}{\includegraphics[angle=0.0]{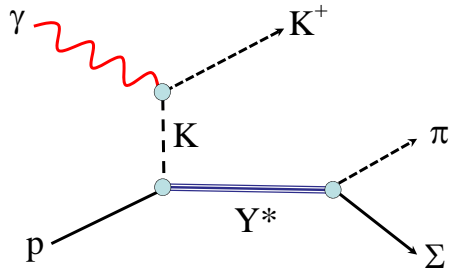}}
  \vspace{-1.0cm}
\end{minipage}\hfill
\begin{minipage}[]{0.38\linewidth}
  \caption{ (Color online) Creation of the three-body $K^+$ $\Sigma$
  $\pi$ final state via an intermediate hyperon in the reaction $\gamma
  + p \to K^+ + \Sigma + \pi$.  In this particular example, a
  $t$-channel exchange enables an off-shell kaon to create a
  $\Lambda(1405)$ that is sub-threshold for on-shell $N\bar{K}$
  reactions. \label{fig:cartoon}}
\end{minipage}
\end{figure}

Satisfactory theoretical understanding of the $\Lambda(1405)$
continues to be a challenge.  For a recent review, see
Ref~\cite{Hyodo:2011ur}. For example, within the spectrum of P-wave
baryons within the relativistic quark model it is a poor quantitative
fit~\cite{Isgur-Karl_PRD18,Capstick-Isgur}.  In models of so-called
chiral unitary state generation among octet baryons and pseudo-scalar
mesons, this state plays an important
role~\cite{Oset-Ramos,Oller:2000fj}. It arises through
channel-coupling of the $N\bar{K}$ system, with threshold at 1432 MeV,
and the $\Sigma\pi$ system into which the $\Lambda(1405)$ must decay.
In the flavor symmetric SU(3) limit, chiral unitary models
show~\cite{Jido:2003cb} that two new octets and a singlet of $J^P =
\frac{1}{2}^-$ states are formed which, when SU(3) is broken, leads to
the presence of two $I=0$ poles in the hyperon sector that reside near
the mass of the $\Lambda(1405)$. Meanwhile, the $I=1$ poles vanish or
are very weak.  Thus, it is of interest to see how well the
``two-pole'' picture of the $\Lambda(1405)$ is supported by the data.
One prediction is that the line shape of the $\Lambda(1405)$ depends
on the reaction channel through which it is excited, since a given
reaction can couple with differing strengths to the two predicted
poles.  In the present case we look using real photoproduction off a
proton target.  This is illustrated in Fig.~\ref{fig:cartoon} which
shows how an off-shell kaon exchange can create the $\Lambda(1405)$
that is sub-threshold for a free kaon plus a nucleon.  A calculation
of this channel was reported in Ref.~\cite{Nacher:1998mi} and compared to our
results in Ref.~\cite{KMRS}.

The CLAS measurements are described in detail in Ref.~\cite{KMRS}.
Briefly, tagged, real, unpolarized photons~\cite{sober} with a highest
energy of 3.8 GeV impinged upon a 40 cm liquid hydrogen target. About
20 billion events with at least two charged tracks in the
large-acceptance toroidal spectrometer~\cite{CLAS-NIM} were recorded,
and in the analysis about $1.41\times10^6$ events with $K^+ \Sigma
\pi$ were isolated.  Charged kaons, protons and pions were identified
by time of flight over a four to five meter flight path.

The $\Sigma^+$ was reconstructed via both of its decay modes, a proton
with an undetected $\pi^0$, or a $\pi^+$ with an undetected neutron.
The $\Sigma^-$ was reconstructed via its decay to $\pi^-$ and an
undetected neutron.  Regions of kinematic overlap of the two charged
$\Sigma$'s were cut away.  The $\Sigma^0\pi^0$ final state was
reconstructed through its decay via $\Lambda \to \pi^- p$ with an
undetected $\pi^0$ and photon.  The undetected particles were
identified via missing mass, and a 1C kinematic fit was employed to
optimize resolution.  The exception was the $\Sigma^0\pi^0$ case where
two neutral particles were missing, so the selection was made using a
fixed range of missing masses.

The two identified backgrounds removed were from $K^+\Sigma^0(1385)$
production and from $K^*\Sigma^0$ production.  The first of these was
measured directly in its dominant decay mode to $\Lambda\pi^0$ in
P-wave.  Using the 11.7\% branching fraction to $\Sigma^\pm\pi^\mp$ we
could accurately subtract this contribution; this was
generally a small fraction of the total $\Sigma\pi$ data.  Coherent
interference with S-wave decays was in principle canceled due to the
full angular coverage of the $\Sigma\pi$ final state in the $Y^*$ rest
frame.  $K^*\Sigma$ production resulted in a broad background
underneath the whole of the $\Sigma\pi$ spectrum, with varying
kinematic overlap in different bins of $W$.  Studies showed that
making increasingly drastic rejection cuts of events in the wide $K^*$
region made no change in the line shapes of the $\Sigma\pi$ results.
We were unable to find any hint of coherent interference due to this
background.  Both backgrounds were modeled in a Monte Carlo simulation
and subtracted away.  Details of the analysis procedures, including
systematic studies, can be found in Refs.~\cite{KMRS}
and~\cite{Moriya-thesis}.

The analysis used 100 MeV wide bins in $\gamma p$ center of
mass energy $W$ to present the results, and it will be seen that the
line shapes differ as $W$ increases.

In this article we present two fits to the $\Sigma\pi$ line shape data
that are different from the ``best'' overall fit described in
Ref.~\cite{KMRS}.  In particular, we first show the fit obtained when
allowing two $I=0$ Breit-Wigner (BW) amplitudes to fit only the
$\Sigma^0\pi^0$ data.  This final state cannot be reached via an $I=1$
decay by reason of the isospin-addition coefficients, and we neglect
$I=2$ processes.  Thus, this fit can serve to test the two-pole
picture of the $\Lambda(1405)$.  Second, we present a fit that adds a
single $I=1$ BW amplitude to the previous two $I=0$ amplitudes and use
this to fit all three $\Sigma\pi$ states at once.  This serves to
characterize the isovector piece of the interaction needed to
represent the experimental data.  This amplitude combination leads
to a result that is statistically less good than the one described in
Ref.~\cite{KMRS} which used \underline{one} $I=0$ amplitude and
\underline{two} $I=1$ amplitudes.  On the other hand, it is closer to
some theoretical biases about the likely two-pole structure of the
$\Lambda(1405)$.  Section~\ref{sec:model} outlines the fit procedure,
and Section~\ref{sec:results} discusses the results.

\section{Model for Mass Distributions}
\label{sec:model}
As described in Ref.~\cite{KMRS}, we posit that the $\Sigma\pi$ part
of the three-body $K^+, \Sigma, \pi$ final states arises via an
interaction schematically labeled $\hat{T}^{(I)}$, where $I$ denotes
the isospin of the $\Sigma\pi$ system.  Let the matrix element squared
be denoted by

\begin{equation}
|t_I|^2 \equiv |\langle I,I_3=0 |\hat T^{(I)}|\gamma p\rangle|^2.
\end{equation}

\noindent The matrix element expressions that give the three measured
final state charge combinations are then given by the coherent sum of
the appropriately Clebsch-Gordon-weighted $t_I$ elements as

\begin{eqnarray}
|T_{\pi^- \Sigma^+}|^2 &=& \frac{1}{3}|t_0|^2 + \frac{1}{2}|t_1|^2  - \frac{2}{\sqrt{6}}|t_0||t_1|\cos{\phi_{01}} \label{eq:sigppim},\\
|T_{\pi^0 \Sigma^0}|^2 &=& \frac{1}{3}|t_0|^2 \label{eq:sig0pi0},\\
|T_{\pi^+ \Sigma^-}|^2 &=& \frac{1}{3}|t_0|^2 + \frac{1}{2}|t_1|^2  + \frac{2}{\sqrt{6}}|t_0||t_1|\cos{\phi_{01}}. \label{eq:sigmpip}
\end{eqnarray}

\noindent At a fixed $\gamma p$ center of mass energy $W$ the matrix
elements are parametrized as relativistic Breit-Wigner functions
$B_I(m)$, where $m$ is the $\Sigma\pi$ mass, with an overall strength
$C(W)$ and a strong production phase $\Delta\phi_I$ as
\begin{equation}
t_I(m) = C_I(W) e^{i \Delta\phi_I} B_I(m). 
\label{eq:normform}
\end{equation}
The Breit-Wigner amplitude form is
\begin{equation} 
B_I(m) = \sqrt{\frac{2}{\pi}} \left[ \frac{\sqrt{m m_R \Gamma_I^0
  \left( \frac{q}{q_R}\right )^{2L}}}{m_R^2 - m^2 -
  i m_R \Gamma_{\mathrm{tot}}(q)} \right]
\label{eq:bwamp}
\end{equation} 
where $m_R$ is the centroid of the resonance distribution,
$\Gamma_I^0$ is the fixed decay width to a given final state, and
$\Gamma_{\mathrm{tot}}(q)$ is the total width to all final states.
The available momentum in the decaying hyperon center of mass system
is $q$, and in this frame $q_{R}$ is the available decay momentum at
$m = m_R$.  Angular momentum $L$ is zero in the present case of an
S-wave decay of the odd-parity $\Lambda(1405)$ decaying to a
pseudo-scalar meson and an octet baryon.

The mass-dependent width in the denominator was treated in a way that
accounts for channel coupling using the Flatt\'e
prescription~\cite{Flatte}.  That is, if decay mode ``1'' is the
$\Sigma\pi$ final state, and decay mode ``2'' is the $N\bar{K}$ final
state, then part of the mass range is below threshold for mode 2.
Nevertheless, mode 2 has influence on mode 1 both below and above
threshold.  To preserve unitarity and the analytic form of the
amplitude, we analytically continuing the center of mass momentum $q$
of decay mode 2 to below its threshold.  We write
\begin{equation}
\Gamma_{\mathrm{tot}}(m) = \Gamma_{I,1}(q_1(m)) + \Gamma_{I,2}(q_2(m)), 
\end{equation}
\noindent
where the decay channels are described by width
\begin{equation} 
\Gamma_{I,j}(q) = \Gamma^0_{I,j} \frac{m_R}{m} \left(\frac{q_j(m)}{q_R}\right)^{2L+1}.
\end{equation} 
\noindent
Below mode 2 threshold $m_{\mathrm{thresh}}$, the momentum $q_2(m)$ is
nominally zero.  However, we continue the momentum to imaginary values
for $m<m_{\mathrm{thresh}}$.  Furthermore, we introduce a Flatt\'e
factor for the branching fraction of the decay modes as
\begin{equation} 
\gamma = \Gamma^0_{I,2}/\Gamma^0_{I,1}.
\end{equation} 
\noindent
This factor is determined by the fits, and indicates the
ratio of the channel coupling.

Finally, the mass distribution is computed as a cross section
differential in $\Sigma\pi$ mass $m$ by including all the flux and
phase space factors for the initial and final states as
\begin{equation}
\frac{d\sigma_{ab}}{dm} = 
\frac{(\hbar c)^2\alpha}{64\pi^3} \frac{p_{K^+}q}{p_{\gamma p} W^2} | T_{\pi^a\Sigma^b}|^2.
\end{equation}
Here $p_{K^+}$ is the c.m. kaon momentum and $p_{\gamma p}$ is the
initial state c.m. momentum.  We have factored out the
electromagnetic coupling strength $\alpha$ so that $C(W)$ becomes an
effective strong coupling in units of $\sqrt{\mathrm{GeV}}$.  This
expression is integrated over all kaon production angles, a step
that was necessary to have adequate statistics.  Similarly, it is
integrated over all $\Sigma$ decay angles in the hyperon rest frame,
as mentioned, to ensure cancellation of any interference with
P-wave decays from the $\Sigma(1385)$ (which is anyway known
to be small and was incoherently subtracted).

\section{Results}
\label{sec:results}

\begin{figure}[htpb]
\centering
\begin{minipage}[]{0.60\linewidth}
  \resizebox{1.0\textwidth}{!}{\includegraphics[angle=0.0]{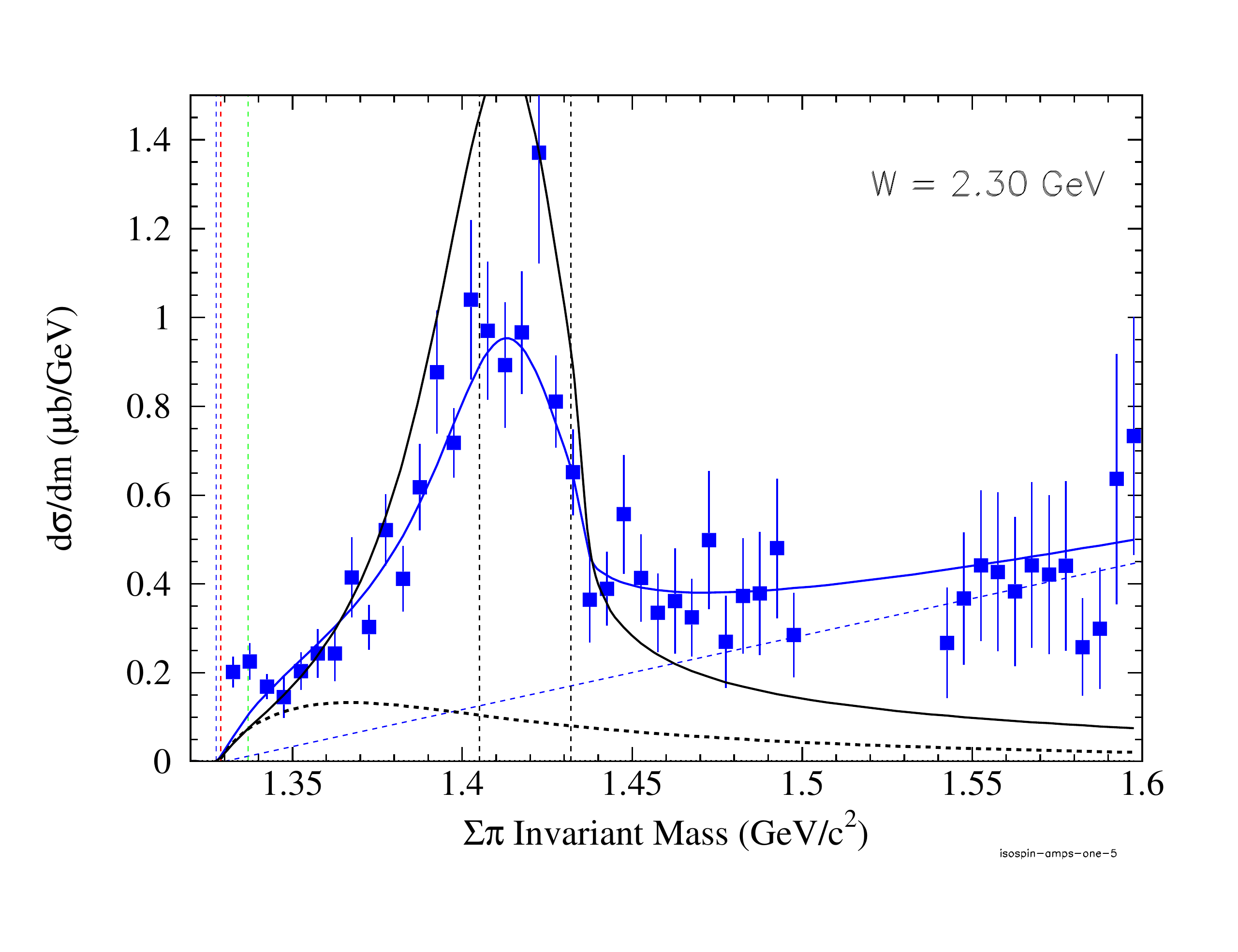}}
  \vspace{-1.0cm}
\end{minipage}\hfill
\begin{minipage}[]{0.40\linewidth}
  \caption{ (Color online) Two $I=0$ amplitudes fitted to
  $\Sigma^0\pi^0$ only (solid and dashed black curves).  The error
  bars are combined statistical and systematic point-to-point
  uncertainties.  The solid blue curve shows the fit to the
  $\Sigma^0\pi^0$ data (blue points).  The three vertical dotted lines
  at the left are thresholds for the three decay modes, and vertical
  black dotted lines mark the nominal 1.405 GeV mass and the location
  of the $N\bar{K}$ threshold $m_{\mathrm{thresh}}$.  The incoherent
  background is shown as a thin dashed line (blue).
  \label{fig:w=2.3_1} }      
\end{minipage}
\end{figure}

\begin{figure}[htpb]
\centering
\begin{minipage}[]{0.85\linewidth}
  \vspace{-1.0cm}
  \resizebox{1.0\textwidth}{!}{\includegraphics[angle=0.0]{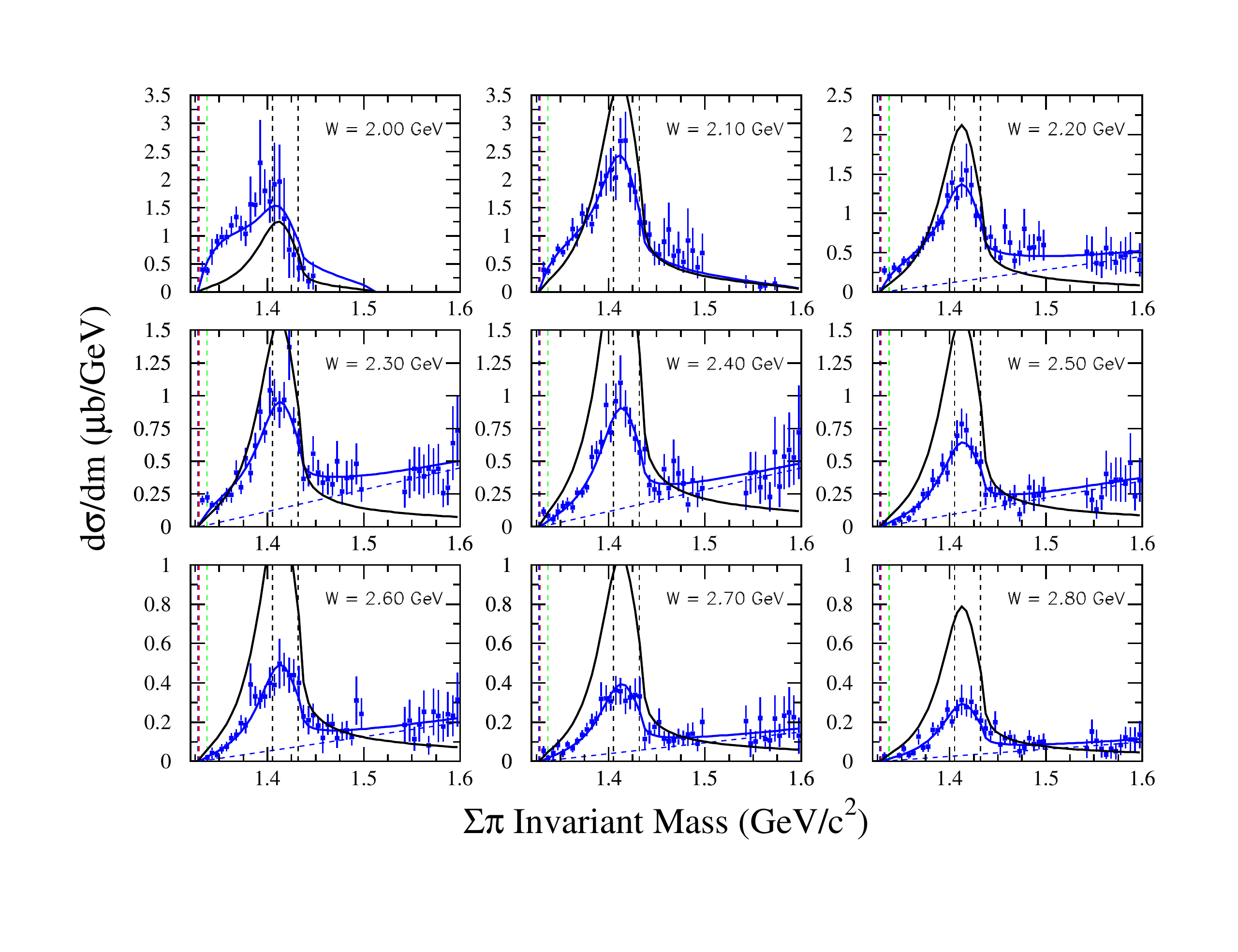}}
\end{minipage}\hfill
\begin{minipage}[]{0.15\linewidth}
  \caption{ (Color online) Two $I=0$ amplitudes fitted to $\Sigma^0\pi^0$
  only. Data and fits for $\Sigma^0 \pi^0$, with each panel showing
  the indicated value of $W$.  Data and curve shapes and colors are as
  in Fig.~\ref{fig:w=2.3_1}.  The curves in all panels use the same
  amplitude parameters from
  Table~\ref{resultstable1}. \label{fig:page13_1}}
\end{minipage}
\end{figure}

\begin{figure}[htpb]
\centering
\begin{minipage}[]{0.60\linewidth}
  \resizebox{1.0\textwidth}{!}{\includegraphics[angle=0.0]{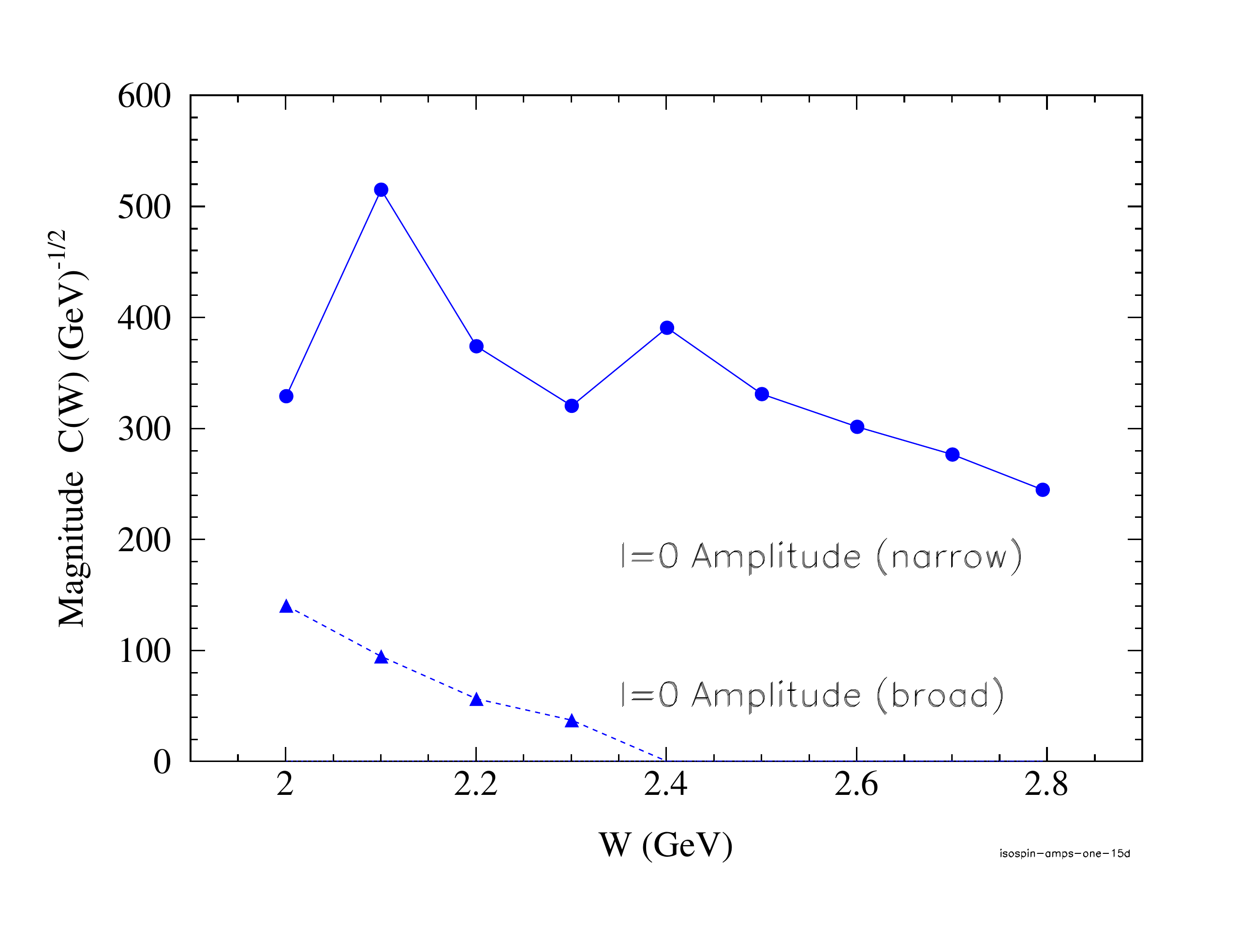}}
  \vspace{-1.0cm}
\end{minipage}\hfill
\begin{minipage}[]{0.40\linewidth}
  \caption{ (Color online) For two $I=0$ amplitudes fitted to
  $\Sigma^0\pi^0$ only, the figure shows the strength of each of the
  isospin amplitudes, $C_0(W)$, as a function of $W$, per
  Eq.~\ref{eq:normform}.
  \label{fig:page15_1} }      
\end{minipage}
\end{figure}

First we consider the fit obtained when using two $I=0$ amplitudes
fitted to the data for $\Sigma^0\pi^0$ only.  This decay mode should
be dominated by the $I=0$ amplitude, assuming $I=2$ can be neglected.
Figure~\ref{fig:w=2.3_1} shows a sample result in one 100 MeV wide bin
of $W$ at 2.3 GeV.  The solid blue curve shows the total fit, and the
solid and dashed black curves show the cross section that would result
from each of the two amplitudes separately.  Also shown is a linear
background (not part of the fit) from threshold to 1.6 GeV that was
included to account for the imperfect modeling of background in the
experimental analysis.  Data points near the location of the
$\Lambda(1520)$ position have also been suppressed.  The fit was done
not just to the $W$ bin shown, but to all nine bins at once.  The full
data set is shown in Fig.~\ref{fig:page13_1}.  In each panel the black
curves have the same shape, differing only in overall weight.  The
weights, the $C_I(W)$, are shown in Fig.~\ref{fig:page15_1}.

It is evident visually that the fit is good; the overall reduced
$\chi^2$ of the fit was 0.89. The results of the fit are given in
Table~\ref{resultstable1}.  The two $I=0$ amplitudes have centroids at
1329 and 1390 MeV, respectively, with respective widths that both
differ when compared to the 50 MeV RPP value for the $\Lambda(1405)$
width.  This is in part explained by the action of the channel
coupling.  It is seen in the table that the Flatt\'e factor for the
narrower amplitude (the only one permitted in the fit) is large, at
1.5.  This has the effect of distorting the line shape around the
$N\bar{K}$ threshold, as seen in the figures, but it also has the
effect of making the most probable mass somewhat above 1405 MeV, even
though the nominal centroid is at 1329.  The goodness of this fit may
be taken as positive evidence for the two-pole solution for the $I=0$
state(s).  Figure~\ref{fig:page15_1} shows how the two components of
the fit vary in strength as $W$ increases: evidently the broader one
fades out quickly while the dominant narrower one remains strong.  We
have no interpretation to offer for this behavior, nor for the
relative production phase between the two amplitudes given in the
table.

\begin{table}[h]
\caption{ Results of the fit using two $I=0$
  Breit-Wigner line shapes fitted to only the $\Sigma^0\pi^0$
  channel. 
  The uncertainties reflect the stability of repeated fits
  with varying initial values. ``N/A'' means no free parameter was allowed.}
\begin{center}
\begin{tabular}{ccccc}
\hline
\hline
Amplitude & Centroid $m_R$ &   Width $\Gamma_0$ & Phase $\Delta\Phi_I$  & Flatt\'e $\gamma$ \\ 
          & (MeV)          &   (MeV)            & (radians)             & Factor            \\ 
\hline 
$I=0$    & $1329\pm 10     $ & $20\pm 10$ & N/A          & $1.5\pm 0.3$\\
$I=0$    & $1390\pm 10     $ & $174\pm20$ & $-0.2\pm0.3$ & N/A         \\
\hline
\hline
\end{tabular}
\end{center}
\label{resultstable1}
\end{table}

\begin{figure}[h]
\centering
\begin{minipage}[]{0.60\linewidth}
  \resizebox{1.00\textwidth}{!}{\includegraphics[angle=0.0]{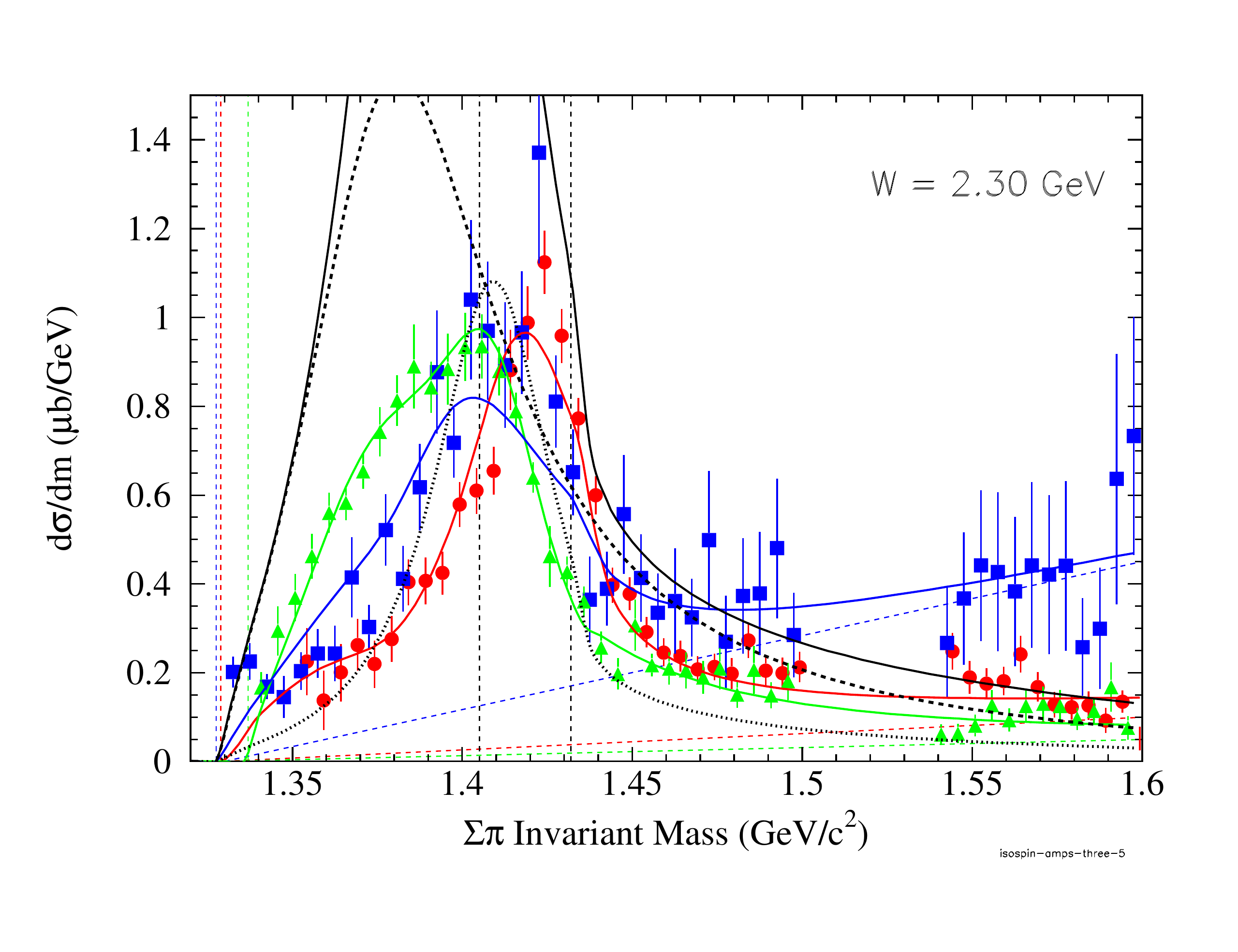}}
  \vspace{-1.0cm}
\end{minipage}\hfill
\begin{minipage}[]{0.40\linewidth}
  \caption{ (Color online) 
  Fits of the isospin amplitudes $I=0$ (solid and dashed black curves)
  and $I=1$ (dotted curve) at the given value of $W$. Data and
  fitted curves for the three charge states are: red curve and circles
  $\Sigma^+ \pi^-$, blue curve and squares: $\Sigma^0 \pi^0$, green
  curve and triangles: $\Sigma^- \pi^+$.  The error bars are combined
  statistical and systematic point-to-point uncertainties. The three
  vertical dotted lines at the left are thresholds for the respective
  decay modes, and vertical dotted lines mark the nominal 1.405 GeV mass
  and the location of the $N\bar{K}$ threshold $m_{\mathrm{thresh}}$.
  The incoherent background is shown as the thin dashed lines.
  \label{fig:w=2.3_3} } 
\end{minipage}
\end{figure}

Next we consider the best fit obtained when using two $I=0$ amplitudes
and a single $I=1$ amplitude fitted to the data for all three charge
final states: $\Sigma^+\pi^-$, $\Sigma^0\pi^0$ and $\Sigma^-\pi^+$.
The fit parameters are given in Table~\ref{resultstable3}; the reduced
$\chi^2$ for this fit was 2.60.  Figure~\ref{fig:w=2.3_3} shows a
sample result in one 100 MeV wide bin of $W$ at 2.3 GeV.  The colored
curves show the total fits to each charge state, and the solid,
dashed, and dotted black curves show the cross sections that would
arise from each of the three amplitudes separately.  The dotted black
curve shows the $I=1$ piece of the reaction.  Its width is similar to
the others (54 MeV), with a centroid at 1367 MeV, and a Flatt\'e
factor that is close to unity, like the one for the $I=0$ piece.
The colored curves show fairly successful modeling of the split
between the three charge combinations.

\begin{table}[h]
\caption{ Results using two $I=0$ and one $I=1$
  Breit-Wigner line shapes and fitting to all final charge states simultaneously.
  The uncertainties reflect the stability of repeated fits
  with varying initial values. ``N/A'' means no free parameter was allowed.}
\begin{center}
\begin{tabular}{ccccc}
\hline
\hline
Amplitude & Centroid $m_R$ &   Width $\Gamma_0$ & Phase $\Delta\Phi_I$  & Flatt\'e $\gamma$ \\ 
          & (MeV)          &   (MeV)            & (radians)             & Factor            \\ 
\hline 
$I=0$ (low mass)  & $1338\pm10$ & $ 44\pm 10$  & N/A               & $0.94\pm 0.20$\\
$I=0$ (high mass) & $1384\pm10$ & $ 76\pm 10$  & $1.8  \pm 0.4 $   & N/A           \\
$I=1$             & $1367\pm20$ & $ 54\pm 10$  & $2.2  \pm 0.4 $   & $1.19\pm 0.20$\\

\hline
\hline
\end{tabular}
\end{center}
\label{resultstable3}
\end{table}

\begin{figure}[h]
\centering
\begin{minipage}[]{0.85\linewidth}
  \resizebox{1.00\textwidth}{!}{\includegraphics[angle=0.0]{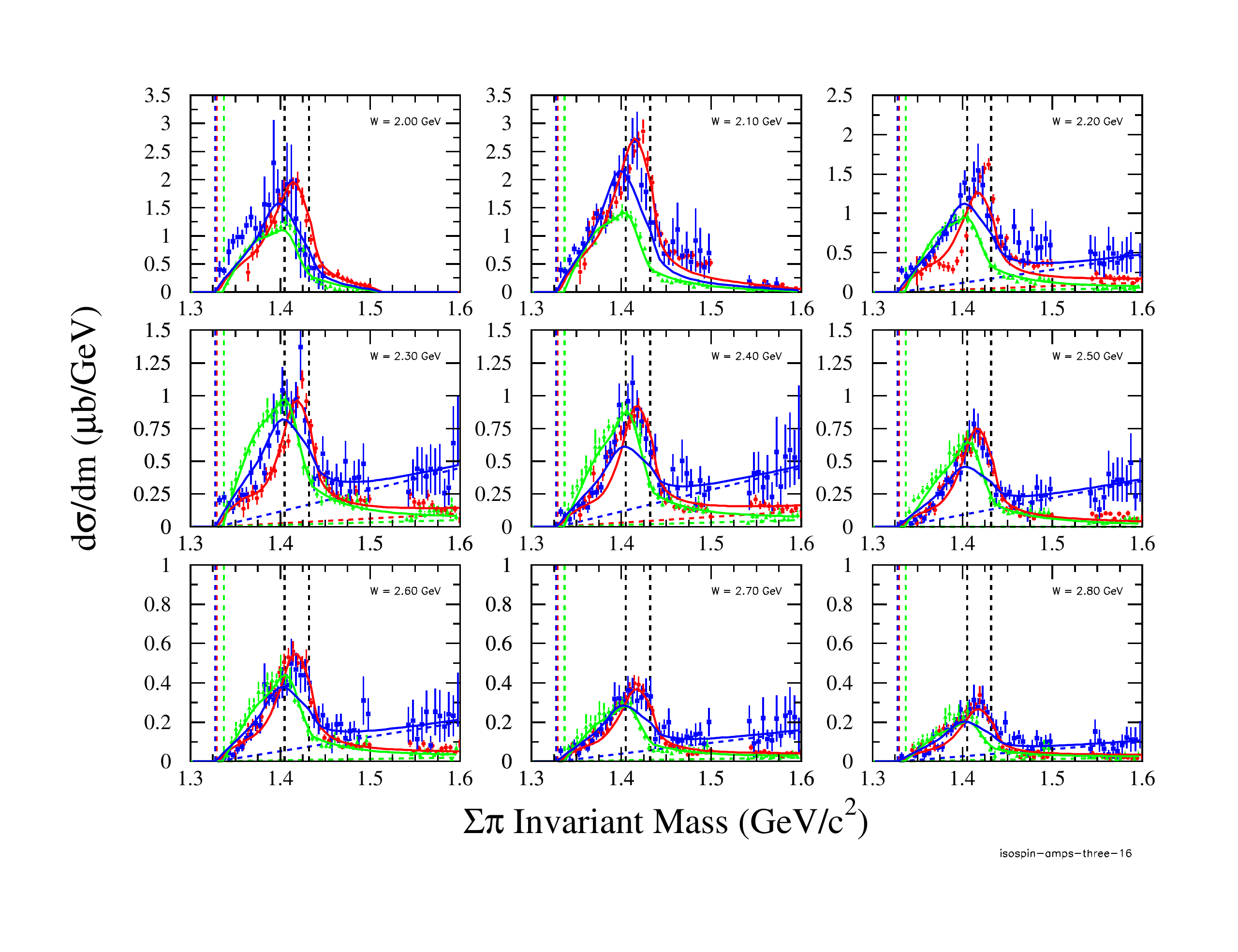}}
  \vspace{-1.0cm}
\end{minipage}\hfill
\begin{minipage}[]{0.15\linewidth}
  \caption{ (Color online) Fits to two $I=0$ amplitudes and one $I=1$
  amplitude, with each panel showing the
  indicated value of $W$, without the underlying pure isospin states
  shown.  Data and curve shapes and colors are as in
  Fig.~\ref{fig:w=2.3_3}.
  \label{fig:page16_3} }      
\end{minipage}
\end{figure}

\begin{figure}[htpb]
\centering
\begin{minipage}[]{0.60\linewidth}
  \resizebox{1.0\textwidth}{!}{\includegraphics[angle=0.0]{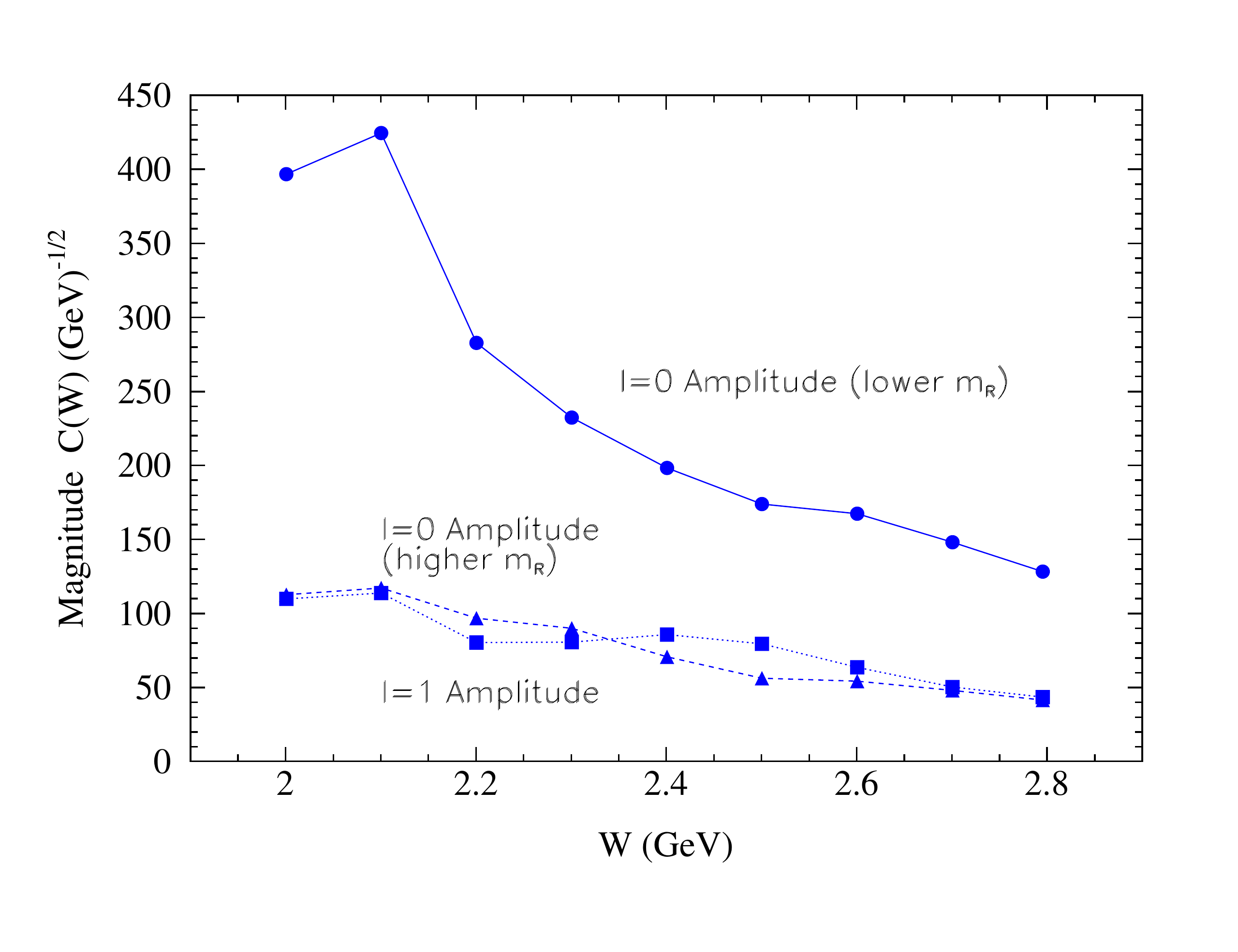}}
  \vspace{-1.0cm}
\end{minipage}\hfill
\begin{minipage}[]{0.40\linewidth}
  \caption{ (Color online) For two $I=0$ amplitudes and one $I=1$ amplitude, 
  the figure shows the strength of each of the
  isospin amplitudes, $C_I(W)$ as a function of $W$, per
  Eq.~\ref{eq:normform}.
  \label{fig:page15_3} }      
\end{minipage}
\end{figure}

Again, the fit was done not just to the $W$ bin shown, but to all nine
bins at once.  The full data set is shown in Fig.~\ref{fig:page16_3}.
In each panel the black curves have the same shapes, differing only in
overall weight.  The way in which the normalizations $C_I(W)$ of the
three amplitudes vary with $W$ is shown in Fig.~\ref{fig:page15_3}.
Note that it is the square of these coefficients that fixes the
relative strength of the contributions. The dominant contribution
comes from the $I=0$ amplitude, as expected.  The $I=1$ amplitude is
weaker in magnitude by a factor of about three, and is about equal in
size to the smaller of the two $I=0$ amplitudes.

It is of interest to compare Table~\ref{resultstable3} with the
previous simpler case shown in Table~\ref{resultstable1}. Upon adding
the interference-causing $I=1$ amplitude, the position and width of
the dominant $I=0$ amplitude rises from 1327 MeV to 1338 MeV; both are
close to $\Sigma\pi$ threshold, and agree within the uncertainties.
The Flatt\'e factor remains in the vicinity of 1.0.  The weaker $I=0$
amplitude stays at the same mass, within uncertainties, and gets
narrower.  Finally, one sees that the $I=1$ amplitude has a centroid
at 1367 MeV with a width of 54 MeV.  It had a fitted Flatt\'e factor
as well, and the fit placed it also close to unity.

Choosing the $I=1$ amplitude to have a resonant Breit-Wigner form was,
of course, an assumption made to create a model.  Alternatively one
might construct a non-resonant scattering amplitude with isovector
character.  But if this analysis is accurate, and if it can be
confirmed that an $I=1$ component in the reaction is reasonably fit
with a Breit-Wigner resonance-like structure, it raises the question
of how to interpret the result.  It may point, for example, to the
existence of a low mass $J^P = \frac{1}{2}^-$ $\Sigma^*$ state as
discussed in Ref.~\cite{Zou:2010tc}.  A full theoretical treatment is
now needed.

\section{Conclusions}
\label{sec:conclusions}

The model of the mass distribution for the $\Sigma\pi$ line shape
introduced in Ref.~\cite{KMRS} has been described briefly.  It has
been applied to recent photoproduction data from CLAS in two
alternative combinations in order to further explore possible
interpretation of the experimental results.

It has been shown that the $I=0$ contribution was qualitatively
consistent with the picture of two poles, modeled as two Breit-Wigner
resonances with large channel coupling to the $N\bar{K}$ final state.
It has been shown, moreover, that an $I=1$ component of the reaction
mechanism can give a reasonable explanation of the variation among
charge states in the $\Sigma\pi$ system seen in photoproduction. This
component was also modeled as a Breit-Wigner line shape with Flatt\'e
coupling to the $N\bar{K}$ final state.  As shown in
Table~\ref{resultstable3}, this $I=1$ strength is centered near 1367
MeV.  If this is the correct way to model this component of the
reaction mechanism, it would lend support to existence of a low mass
$\Sigma^*$ state with $J^P=\frac{1}{2}^-$.

Nevertheless, it must be pointed out that a better fit (in the sense
of both $\chi^2$ and the visual appearance of the fit results) was
found~\cite{KMRS} previously with an alternative choice of one $I=0$
amplitude and two $I=1$ amplitudes.  The $\chi^2$ in that study was
2.15.  Neither is the model unique, nor is the data precise enough to
allow only a unique result from the given model.  Clearly, more
modeling and more theoretical work on the interpretation should
follow.
 
\clearpage




\bibliographystyle{model1-num-names}
\bibliography{HYP11_Schumacher}

\begin{thebibliography}{20}
\expandafter\ifx\csname natexlab\endcsname\relax\def\natexlab#1{#1}\fi
\providecommand{\bibinfo}[2]{#2}
\ifx\xfnm\relax \def\xfnm[#1]{\unskip,\space#1}\fi
\bibitem[{Moriya et~al.(2012)Moriya, Schumacher et~al.}]{KMRS}
\bibinfo{author}{K.~Moriya}, \bibinfo{author}{R.~A. Schumacher}, et~al.,
\newblock \bibinfo{title}{{Measurement of the $\Sigma\pi$ Photoproduction Line
  Shapes Near the $\Lambda(1405)$}},
\newblock \bibinfo{journal}{accepted by Phys. Rev. C}  (\bibinfo{year}{2012}).
\bibitem[{Hemingway(1985)}]{Hemingway}
\bibinfo{author}{R.~J. Hemingway},
\newblock \bibinfo{title}{{Production of $\Lambda(1405)$ in $K^- p$ Reactions
  at 4.2 $\mathrm{GeV}/c$}},
\newblock \bibinfo{journal}{Nucl. Phys.} \bibinfo{volume}{B253}
  (\bibinfo{year}{1985}) \bibinfo{pages}{742}.
\bibitem[{Thomas et~al.(1973)Thomas, Engler, Fisk, and Kraemer}]{Thomas}
\bibinfo{author}{D.~W. Thomas}, \bibinfo{author}{A.~Engler},
  \bibinfo{author}{H.~E. Fisk}, \bibinfo{author}{R.~W. Kraemer},
\newblock \bibinfo{title}{{Strange Particle Production from $\pi^- p$
  Interactions at 1.69 $\mathrm{GeV}/c$}},
\newblock \bibinfo{journal}{Nucl. Phys.} \bibinfo{volume}{B56}
  (\bibinfo{year}{1973}) \bibinfo{pages}{15--45}.
\bibitem[{Braun et~al.(1977)Braun, Grimm, Hepp, Strobele, Thol
  et~al.}]{Braun:1977wd}
\bibinfo{author}{O.~Braun}, \bibinfo{author}{H.~Grimm},
  \bibinfo{author}{V.~Hepp}, \bibinfo{author}{H.~Strobele},
  \bibinfo{author}{C.~Thol}, et~al.,
\newblock \bibinfo{title}{{New Information About the Kaon-Nucleon-Hyperon
  Coupling Constants: $g(\bar{K} n \Sigma(1197))$, $g(\bar{K} n \Sigma(1385))$
  and $g(\bar{K} n \Lambda(1405))$}},
\newblock \bibinfo{journal}{Nucl.Phys.} \bibinfo{volume}{B129}
  (\bibinfo{year}{1977}) \bibinfo{pages}{1}.
\bibitem[{Ahn et~al.(2003)}]{Ahn}
\bibinfo{author}{J.~K. Ahn}, et~al.,
\newblock \bibinfo{title}{{$\Lambda(1405)$ photoproduction at SPring-8/LEPS}},
\newblock \bibinfo{journal}{Nucl. Phys.} \bibinfo{volume}{A721}
  (\bibinfo{year}{2003}) \bibinfo{pages}{715--718}.
\bibitem[{Niiyama et~al.(2008)}]{Niiyama}
\bibinfo{author}{M.~Niiyama}, et~al.,
\newblock \bibinfo{title}{{Photoproduction of $\Lambda(1405)$ and
  $\Sigma^{0}(1385)$ on the proton at $E_\gamma = 1.5$-$2.4$ GeV}},
\newblock \bibinfo{journal}{Phys. Rev.} \bibinfo{volume}{C78}
  (\bibinfo{year}{2008}) \bibinfo{pages}{035202}.
\bibitem[{Zychor et~al.(2008)}]{Zychor}
\bibinfo{author}{I.~Zychor}, et~al.,
\newblock \bibinfo{title}{{Shape of the $\Lambda(1405)$ hyperon measured
  through its $\Sigma^0 \pi^0$ decay}},
\newblock \bibinfo{journal}{Phys. Lett.} \bibinfo{volume}{B660}
  (\bibinfo{year}{2008}) \bibinfo{pages}{167--171}.
\bibitem[{Agakishiev et~al.(2012)Agakishiev, Balanda, Belver, Belyaev,
  Berger-Chen et~al.}]{Agakishiev:2012xk}
\bibinfo{author}{G.~Agakishiev}, \bibinfo{author}{A.~Balanda},
  \bibinfo{author}{D.~Belver}, \bibinfo{author}{A.~Belyaev},
  \bibinfo{author}{J.~Berger-Chen}, et~al.,
\newblock \bibinfo{title}{{Baryonic resonances close to the $\bar{K}N$
  threshold: the case of $\Lambda(1405)$ in $pp$ collisions}},
\newblock \bibinfo{journal}{Phys.Rev.} \bibinfo{volume}{C85}
  (\bibinfo{year}{2012}) \bibinfo{pages}{035203}.
\bibitem[{Hyodo and Jido(2012)}]{Hyodo:2011ur}
\bibinfo{author}{T.~Hyodo}, \bibinfo{author}{D.~Jido},
\newblock \bibinfo{title}{{The nature of the $\Lambda(1405)$ resonance in
  chiral dynamics}},
\newblock \bibinfo{journal}{Prog.Part.Nucl.Phys.} \bibinfo{volume}{67}
  (\bibinfo{year}{2012}) \bibinfo{pages}{55--98}.
\bibitem[{Isgur and Karl(1978)}]{Isgur-Karl_PRD18}
\bibinfo{author}{N.~Isgur}, \bibinfo{author}{G.~Karl},
\newblock \bibinfo{title}{{P Wave Baryons in the Quark Model}},
\newblock \bibinfo{journal}{Phys. Rev.} \bibinfo{volume}{D18}
  (\bibinfo{year}{1978}) \bibinfo{pages}{4187}.
\bibitem[{Capstick and Isgur(1986)}]{Capstick-Isgur}
\bibinfo{author}{S.~Capstick}, \bibinfo{author}{N.~Isgur},
\newblock \bibinfo{title}{{Baryons in a Relativized Quark Model with
  Chromodynamics}},
\newblock \bibinfo{journal}{Phys. Rev.} \bibinfo{volume}{D34}
  (\bibinfo{year}{1986}) \bibinfo{pages}{2809}.
\bibitem[{Oset and Ramos(1998)}]{Oset-Ramos}
\bibinfo{author}{E.~Oset}, \bibinfo{author}{A.~Ramos},
\newblock \bibinfo{title}{{Non perturbative chiral approach to s-wave $\bar{K}
  N$ interactions}},
\newblock \bibinfo{journal}{Nucl. Phys.} \bibinfo{volume}{A635}
  (\bibinfo{year}{1998}) \bibinfo{pages}{99--120}.
\bibitem[{Oller and Meissner(2001)}]{Oller:2000fj}
\bibinfo{author}{J.~A. Oller}, \bibinfo{author}{U.~G. Meissner},
\newblock \bibinfo{title}{{Chiral dynamics in the presence of bound states:
  Kaon nucleon interactions revisited}},
\newblock \bibinfo{journal}{Phys. Lett.} \bibinfo{volume}{B500}
  (\bibinfo{year}{2001}) \bibinfo{pages}{263--272}.
\bibitem[{Jido et~al.(2003)Jido, Oller, Oset, Ramos, and
  Meissner}]{Jido:2003cb}
\bibinfo{author}{D.~Jido}, \bibinfo{author}{J.~A. Oller},
  \bibinfo{author}{E.~Oset}, \bibinfo{author}{A.~Ramos}, \bibinfo{author}{U.~G.
  Meissner},
\newblock \bibinfo{title}{{Chiral dynamics of the two $\Lambda(1405)$ states}},
\newblock \bibinfo{journal}{Nucl. Phys.} \bibinfo{volume}{A725}
  (\bibinfo{year}{2003}) \bibinfo{pages}{181--200}.
\bibitem[{Nacher et~al.(1999)Nacher, Oset, Toki, and Ramos}]{Nacher:1998mi}
\bibinfo{author}{J.~C. Nacher}, \bibinfo{author}{E.~Oset},
  \bibinfo{author}{H.~Toki}, \bibinfo{author}{A.~Ramos},
\newblock \bibinfo{title}{{Photoproduction of the $\Lambda(1405)$ on the proton
  and nuclei}},
\newblock \bibinfo{journal}{Phys. Lett.} \bibinfo{volume}{B455}
  (\bibinfo{year}{1999}) \bibinfo{pages}{55--61}.
\bibitem[{Sober et~al.(2000)}]{sober}
\bibinfo{author}{D.~I. Sober}, et~al.,
\newblock \bibinfo{title}{{The bremsstrahlung tagged photon beam in Hall B at
  JLab}},
\newblock \bibinfo{journal}{Nucl. Instrum. Meth.} \bibinfo{volume}{A440}
  (\bibinfo{year}{2000}) \bibinfo{pages}{263--284}.
\bibitem[{Mecking et~al.(2003)}]{CLAS-NIM}
\bibinfo{author}{B.~A. Mecking}, et~al.,
\newblock \bibinfo{title}{{The CEBAF Large Acceptance Spectrometer (CLAS)}},
\newblock \bibinfo{journal}{Nucl. Instrum. Meth.} \bibinfo{volume}{A503}
  (\bibinfo{year}{2003}) \bibinfo{pages}{513--553}.
\bibitem[{Moriya(2010)}]{Moriya-thesis}
\bibinfo{author}{K.~Moriya},
\newblock \bibinfo{title}{{Measurement of the Lineshape, Differential
  Photoproduction Cross Section, Spin and Parity of the $\Lambda(1405)$ Using
  CLAS at Jefferson Lab}},
\newblock \bibinfo{journal}{Thesis, Carnegie Mellon University}
  (\bibinfo{year}{2010}). \bibinfo{note}{Available online at {\tt
  http://www.jlab.org/Hall-B/general/clas\_thesis.html}}.
\bibitem[{Flatt\'e(1976)}]{Flatte}
\bibinfo{author}{S.~M. Flatt\'e},
\newblock \bibinfo{title}{{Coupled - Channel Analysis of the $\pi \eta$ and $K
  \bar{K}$ Systems Near $K \bar{K}$ Threshold}},
\newblock \bibinfo{journal}{Phys. Lett.} \bibinfo{volume}{B63}
  (\bibinfo{year}{1976}) \bibinfo{pages}{224}.
\bibitem[{Zou(2010)}]{Zou:2010tc}
\bibinfo{author}{B.-S. Zou},
\newblock \bibinfo{title}{{Five-quark components in baryons}},
\newblock \bibinfo{journal}{Nucl. Phys.} \bibinfo{volume}{A835}
  (\bibinfo{year}{2010}) \bibinfo{pages}{199--206}.

\end{thebibliography}



\end{document}